\documentclass [12pt] {article}
 \usepackage{epsfig}
\begin {document}

\title{$J/\Psi$ and Drell-Yan pair production on nuclear targets}

\author{Ya. A. Berdnikov$^1$ \and
        M. M. Ryzhinskiy$^1$\thanks{E-mail: mryzhinskiy@phmf.spbstu.ru} \and
        Yu. M. Shabelski$^2$\thanks{E-mail: shabelsk@thd.pnpi.spb.ru}}
\date{$^1$St.-Petersburg State Polytechnic University, St.-Petersburg, Russia \\
$^2$Petersburg Nuclear Physics Institute, Gatchina, St.-Petersburg, Russia}

\maketitle

 \begin{abstract}
 We estimate the energy losses in the cases of $J/\Psi$ and $l^+l^-$ pair
 production on nuclear targets in terms of effective change of the initial
 beam energy. Our phenomenological results are in reasonable agreement with
 Theoretical calculations.
 \end{abstract}

 \graphicspath{{./eps/}{./}}

 \newpage
 \pagestyle{plain}
 \section{Introduction}
 \vskip 0.5 truecm

 There exist significant nuclear effects in the cases of $J/\Psi$, and even of
 Drell-Yan pair production. These effects are discussed in many papers
 (see below). We present the possibility to estimate them directly from
 the experimental data.

Let us consider the variable
 \begin{equation}
 \label{eq:x0}
 x_0=\frac{p}{p_0},
 \end{equation}
 where $p$ is the momentum
 of produced secondary and $p_0$ is the initial momentum of the beam particle
 both in c.m. frame.
 At high energies in the case of nucleon target the maximum value of
 $x_0$ is close to unity ($(x_0)_{\rm max}\to1$). In the case of nuclear target
 the situation is more complicated
 because there are many different contributions. First of all there exists
 the so called cumulative production (with $x_0>1$)
 \cite{Baldin1,Bayukov1,Frankfurt1}. However it is
 a special process which we are not going to consider in this paper.

 Many other processes result in $(x_0)_{\rm max}<1$ due to
 the initial- and/or final-state interactions \cite{Vogt1}.
 For example in the case of Drell-Yan production there exist initial-state
 effect such as energy loss of the incident quark in nuclear target as well as
 nuclear shadowing \cite{Vasiliev1,Kop1}. In the case of $J/\Psi$ production there
 is an additional source of suppression connected with final-state interactions
 \cite{Kop2}.

 Moreover, in all processes on nuclear targets (except of coherent ones)
 some fraction of energy is used for nuclear disintegration.
 The nuclear target is destroyed and several nucleons, as well as light
nuclear fragments, (say, $\alpha$-particles) appear in the final state.
It can be considered as a phase space limitation. This fraction
 is numerically not so small (it is many times more than the nuclear
 binding energy) \cite{Azimov1,Azimov2}. This source of energy loss also
leads to decrease of $(x_0)_{\rm max}$ but it was not taken into account
in all modern papers.

 In the present paper we will not consider every source of suppression
 separately, but we are going to consider them all together including
 the last one.

 It is clear that all initial-state energy losses are equivalent to
 decrease of the incident beam momentum $p_0$. As the phase space effects
 as well as final state interactions decrease $(x_0)_{max}$, we assume
 that their influence in the considered processes can be effectively
 described by the same way, as the additional decrease of $p_0$. So we
 change $p_0$ by $p_0 - p_A$, where $p_A$ is a phenomenological parameter
 which accounts for all energy losses.

 In what follows we will consider the $A$-dependences of $J/\Psi$ and
 Drell-Yan pair production in terms of $x_{\rm F}$:

 \begin{equation}
 \label{eq:xF}
 x_{\rm F} = \frac{p}{p_0 - p_A}
 \end{equation}
and we assume that it is possible to find the shift $p_A$ from the
condition that the ratio of multiplicities on nucleon and nuclear
targets
\begin{equation}
R_{hA/hp}(x_F) = const (x_F) \simeq 1 ,
\end{equation}
whereas the same multiplicity ratio in terms of $x_0$ has evident
$A$-dependence
\begin{equation}
R_{hA/hp}(x_0) = f(x_0).
\end{equation}

Evidently, such rescale is reasonable only for not very small $x_0$
values.

We will determine the shift $p_A$ from the
experimental data and we will compare them with several independent
 estimations.
 Such approach allows us, in particular, to investigate the energy dependence
 of all nuclear effects.
 In conclusion we will compare our results with theoretical
 calculations \cite{Kop1,Kop2}.

 \section{$A$-dependence of $J/\Psi$ production at large $x_{\rm F}$}
 Charmonium production off nuclei has drawn much attention during the last
 two decades, since the NA3 experiment at CERN \cite{Badier1} has found a
 steep increase of nuclear suppression with rising $x_0$.
 This effect was confirmed later in the same energy range \cite{Katsanevas1}
 as well as at high energies \cite{Leitch1,Alde1,Kowitt1}.

 For the purpose of our analysis we used the experimental data on $J/\Psi$ production
 in proton-nucleus ($pA$) collisions published in \cite{Leitch1,Kowitt1,Gribushin1}.
 We also used data on $J/\Psi$ production in $\pi^{-}A$ collisions
 \cite{Katsanevas1, Abolins1}. We analysed the ratio $R_{A_1/A_2}$ of
 inclusive differential cross sections for $J/\Psi$ hadroproduction on $A_1$ nucleus
 to that on $A_2$:

 \begin{equation}
 \label{eq:ratio}
 R_{A_1/A_2} = \frac{\frac{1}{A_1}\left(\frac{{\rm d}\sigma}{{\rm d}x_0}\right)_{hA_1\rightarrow J/\Psi}}
 {\frac{1}{A_2}\left(\frac{{\rm d}\sigma}{{\rm d}x_0}\right)_{hA_2\rightarrow J/\Psi}}.
 \end{equation}

 The analysis was perfomed in the following way. Suppose one has two
 $x_0$-spectra for $J/\Psi$ production on nucleon and nuclear targets
 or on two different nuclei (light and
 heavy ones). Then one can shift the spectrum that corresponds to the heavy
 nucleus according to Eq. (\ref{eq:xF}) by changing $p_A$ parameter.
 Assuming that nuclear effects are small
 for very light nuclei, it is possible then to calculate the fraction of beam energy/momentum
 spent on nuclear effects. This may be done by
 calculating the ratio of the shifted spectrum to the spectrum that corresponds to the
 light nucleus. When this ratio is close to unity then the corresponding shift
 will give the absolute value of energy/momentum loss caused by the mentioned
 nuclear effects.

\begin{figure}[htb]
 \centering
 \includegraphics[width=.9\hsize]{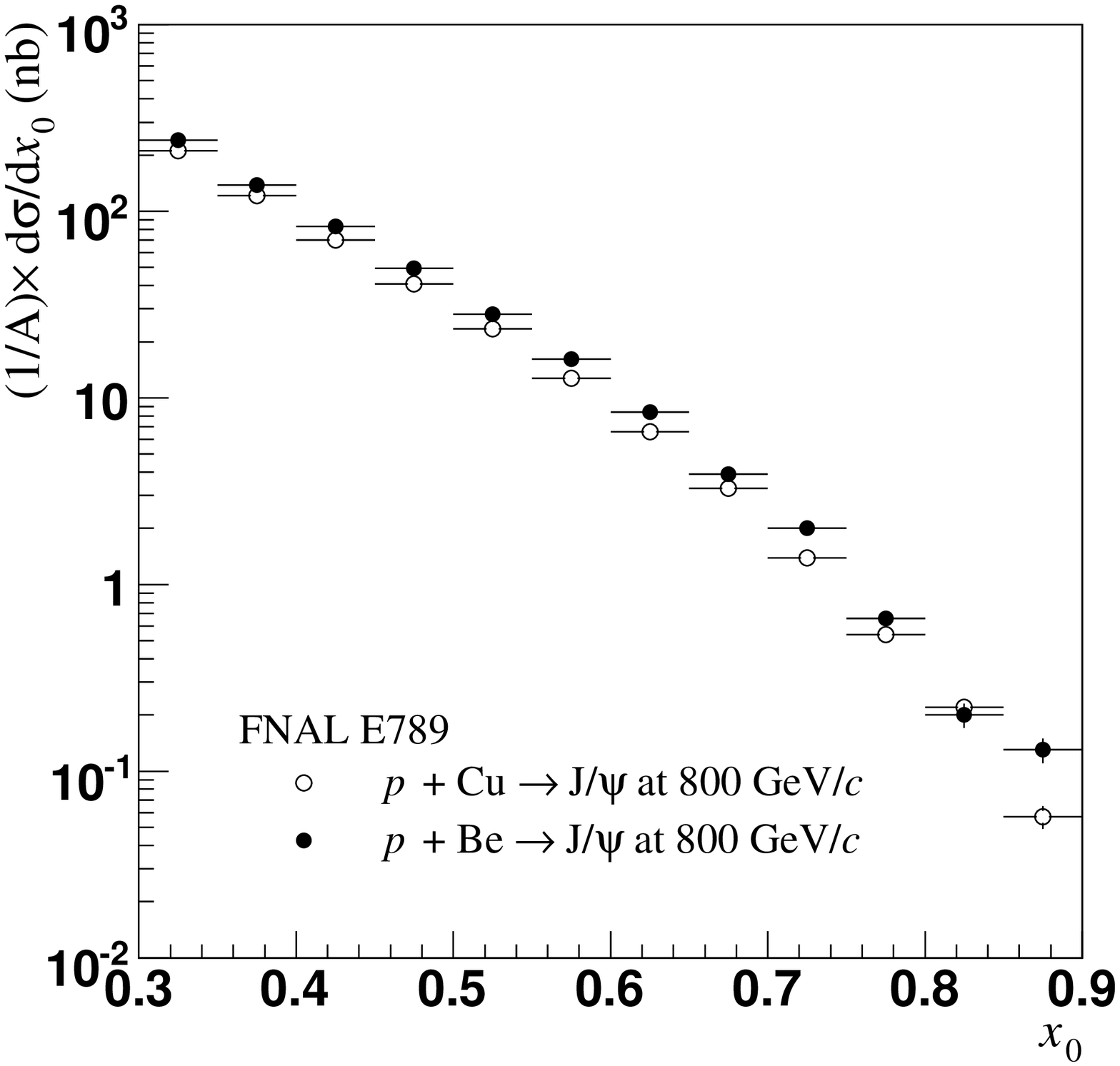}
 \caption{Differential cross section for $J/\Psi$ production in
 $p$Cu and $p$Be collisions at 800~GeV/{\it c} \cite{Kowitt1}.}
 \label{fig:E789initial}
 \end{figure}

 Let us analyse the spectra \cite{Kowitt1} presented in Fig.~\ref{fig:E789initial}.
 The spectra represent
 the differential cross sections for $J/\Psi$'s produced inclusively in 800~GeV/{\it c}
 $p$Cu and $p$Be collisions measured by E789 Collaboration. One can see the
 evident nuclear suppression for copper target.

\begin{figure}[htb]
 \centering
 \includegraphics[width=.9\hsize]{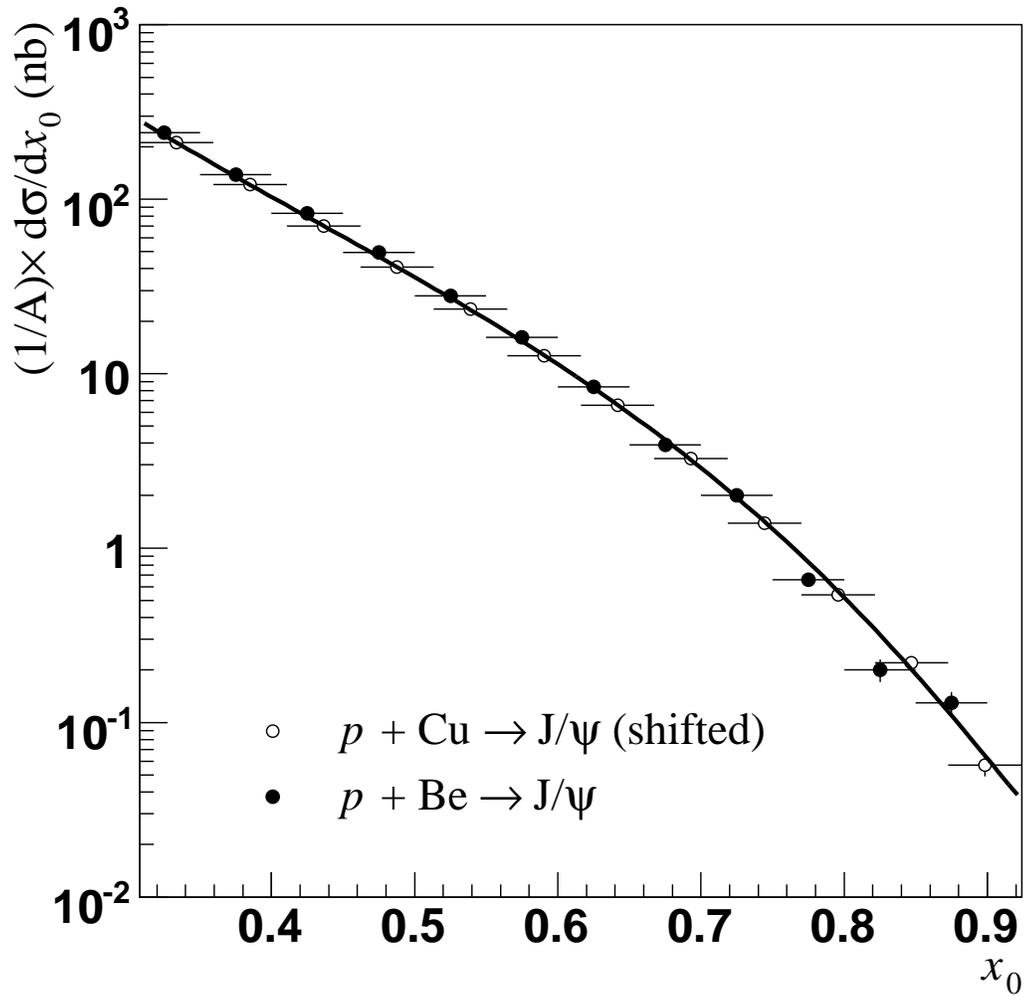}
 \caption{The same as in Fig.~\ref{fig:E789initial}, but the spectrum for Cu was
 shifted according to Eq.~(\ref{eq:xF}) with $p_A=0.5$~GeV/{\it c}.}
 \label{fig:E789shifted}
 \end{figure}

 Now to calculate how much energy/momentum of the beam particle is spent on nuclear
 effects we should shift the spectrum for the heavy nucleus (namely copper). The shifted spectrum
 is presented on Fig.~\ref{fig:E789shifted}. The solid curve in Fig.~\ref{fig:E789shifted}
 represents the fit to the shifted spectrum
 needed for further calculations of the ratio of the presented spectra.

\begin{figure}[htb]
 \centering
 \includegraphics[width=.9\hsize]{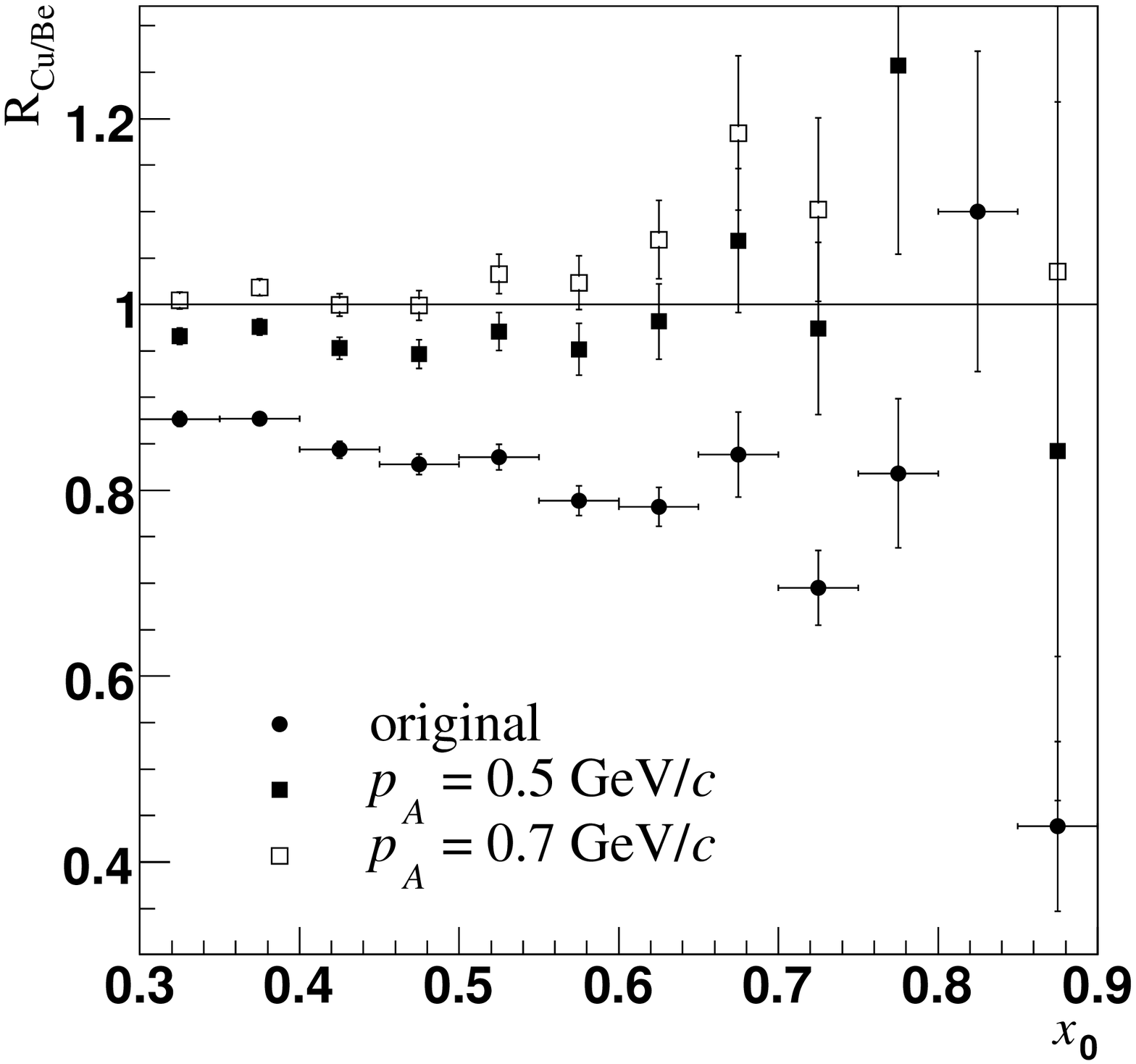}
 \caption{The ratios of inclusive differential cross sections, calculated for different $p_A$
 values.}
 \label{fig:E789ratios}
 \end{figure}

 The calculated ratios Eq.~(\ref{eq:ratio}) one can see in Fig.~\ref{fig:E789ratios}.
 Solid circles correspond to the
 original ratio measured by E789 Collaboration, two other ratios were calculated for different
 shifts: $p_A=0.5$~GeV/{\it c} (solid squares) and $p_A=0.7$~GeV/{\it c} (open squares). These
 $p_A$ values represent the amount of absolute momentum spent on nuclear effects.

 The values of $p_A$ in c.m. frame correspond to the shift

 \begin{equation}
 \label{eq:shift_cm}
 \Delta x^{\rm c.m.}_{\rm F} = \frac{p_A}{p_0}.
 \end{equation}
 For large $x_{\rm F}$ $\Delta x^{\rm lab.}_{\rm F}\approx\Delta x^{\rm c.m.}_{\rm F}$,
 so we can calculate the
 absolute value of energy losses in lab. frame. In our case the shift
 $p_A = 0.5$~GeV/{\it c} corresponds in the lab. frame to the energy losses
 $\Delta p^{\rm lab.}_{\rm Cu/Be}\approx 20$~GeV/{\it c}.

 Unfortunately E789 is the only experiment that measured absolute cross sections for $J/\Psi$
 production in $pA$ collisions for both light and heavy nuclei.
 Experimentally it is much easier to measure ratios of
 the cross sections at once, thus most collaborations present only the ratios
 (without absolute values of the cross sections) as their results. Consequently
 it is difficult to analyse those experiments in such a way we did above.

\begin{figure}[htb]
 \centering
 \includegraphics[width=1.\hsize]{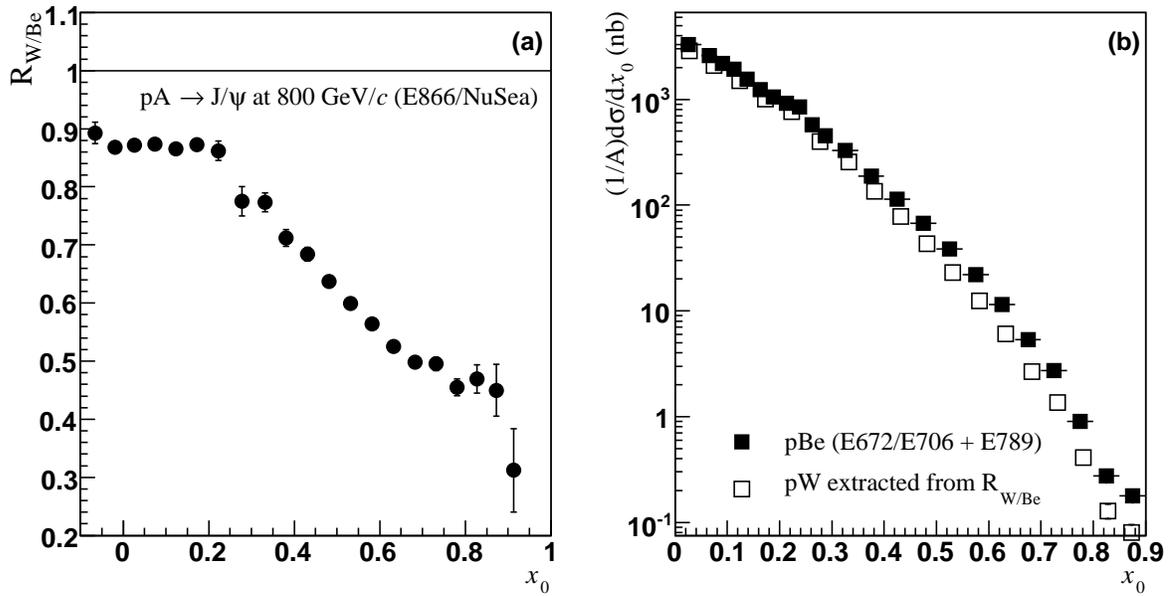}
 \caption{(a) The ratio of inclusive differential cross section for $J/\Psi$ production
 on W target to that on Be measured by E866 \cite{Alde1}.
 (b) Combined spectrum (see text) for Be target (solid squares),
 and spectrum for W (open squares) target extracted from the ratio shown in Fig.~\ref{fig:E866initial}a.}
 \label{fig:E866initial}
 \end{figure}

\begin{figure}[htb]
 \centering
 \includegraphics[width=1.\hsize]{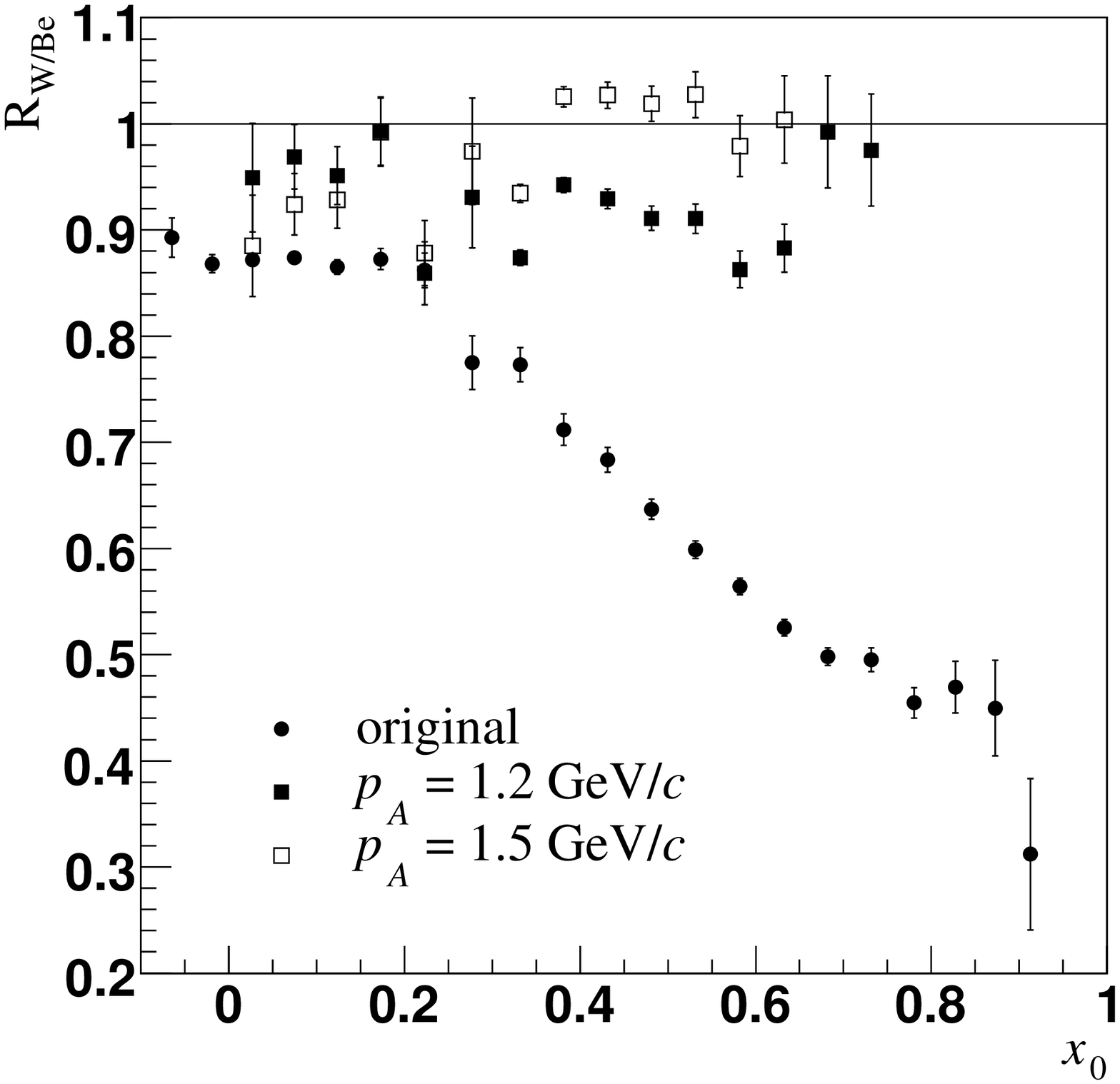}
 \caption{The ratios of inclusive differential cross sections, calculated for different $p_A$
 values.}
 \label{fig:E866ratios}
 \end{figure}

 However the most recent and most precise experiment on $J/\Psi$ production by E866
 Collaboration \cite{Leitch1} was performed at the same energy
 (800 GeV) as E789 and used Be and W targets. Thus using the
 E789 $x_0$-spectrum for beryllium target and E866 ratio of spectrum on beryllium to
 that on tungsten one can extract the absolute $J/\Psi$ production
 cross section for tungsten. Since the $x_0$-scale covered by E866 experiment
 ($-0.1<x_0<0.93$) is larger than one covered by E789 ($0.3<x_0<0.95$), then
 to extract E866 spectrum for W target we
 combined the E789 data with the data obtained by
 E672 and E706 Collaborations for Be target in the range of
 $0.0 < x_0 < 0.5$ \cite{Gribushin1}.
 Fig.~\ref{fig:E866initial} represents the ratio measured by E866
 (Fig.~\ref{fig:E866initial}a) and the absolute spectra for Be target
 (solid squares)
 combined from the mentioned data sets, and for W target
 (open squares) extracted
 from the ratio (Fig.~\ref{fig:E866initial}b).
 Now it is possible to analyse the last two spectra in the same way as was
 done in the case of E789 data (see Fig.~\ref{fig:E789shifted}).
 We omit intermediate calculations and present the final result.
 The shifted ratios at
 two different shift values ($p_A=1.2$~GeV/{\it c} --- solid squares,
 and $p_A=1.5$~GeV/{\it c} --- open squares) are presented in
 Fig.~\ref{fig:E866ratios}. This corresponds to the absolute energy losses
 in the lab. frame $\Delta p^{\rm lab.}_{\rm W/Be}\approx 50$~GeV/{\it c}.

 One can see that the absolute value
 of energy/momentum loss in tungsten is more than two times larger than one
 in copper target, which is rather clear. Indeed, W target is $\approx2.9$
 times heavier than Cu. Consequently, nuclear effects in the former
 should be stronger, thus the value of energy/momentum lost by the projectile
 in W target should be larger than that in Cu.

\begin{figure}[htb]
 \centering
 \includegraphics[width=1.\hsize]{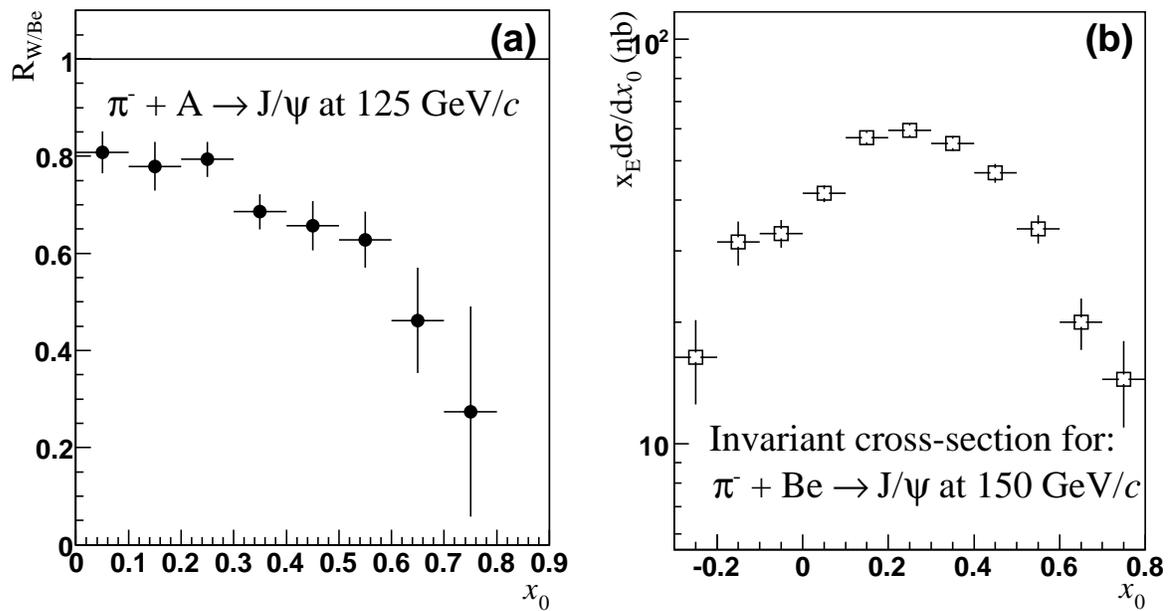}
 \caption{$x_0$-spectra for $J/\Psi$'s produced in $\pi^{-}A$ collisions.
 (a) The ratios of inclusive differential cross sections for $J/\Psi$
 production on tungsten to that on beryllium \cite{Katsanevas1}.
 (b) The invariant differential cross section for $J/\Psi$ production
 on Be target.}
 \label{fig:Pi_initial}
 \end{figure}

\begin{figure}[htb]
 \centering
 \includegraphics[width=1.\hsize]{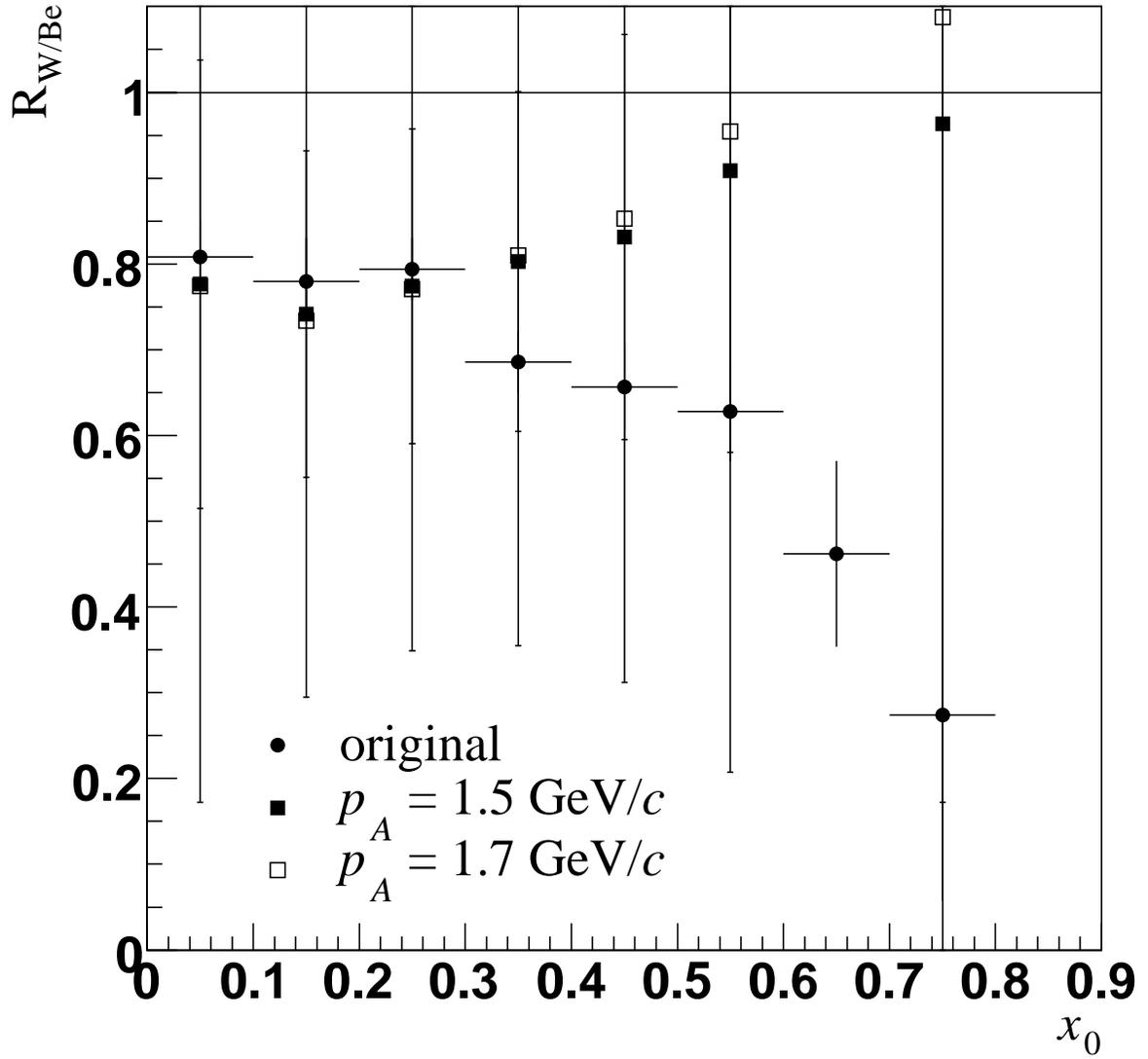}
 \caption{The ratios of inclusive differential cross sections for $J/\Psi$
 production in $\pi^-A$ collisions at 125~GeV/{\it c}, calculated for
 different $p_A$ values.}
 \label{fig:Pi_ratios}
 \end{figure}

 Besides proton induced reactions we considered the data on $J/\Psi$ production
 in $\pi^{-}A$ collisions. The data available are the ratio of differential cross sections
 for $J/\Psi$ production on W to that on Be target at 125~GeV/{\it c}
 \cite{Katsanevas1} (Fig.~\ref{fig:Pi_initial}a),
 and absolute differential cross section for $J/\Psi$ production
 on Be target at 150~GeV/{\it c} \cite{Abolins1}.
 Since the energies for the
 two data sets slightly differ from each other, we build the
 corresponding invariant cross section
 from that in Ref.~\cite{Abolins1}, which is presented in Fig.~\ref{fig:Pi_initial}b.
 To extract the spectrum for a heavy nucleus (namely, tungsten) we applied the procedure
 described above. Omitting intermediate results we present the range of $p_A$
 values obtained from the analysis (Fig.~\ref{fig:Pi_ratios}):

 \begin{equation}
 p_A = 1.5 - 1.7 {\rm ~GeV}/c,
 \nonumber
 \end{equation}
 which correspond to $\Delta p^{\rm lab.}_{\rm W/Be} \approx 12$~GeV/c.
 The obtained $\Delta p^{\rm lab.}_{\rm W/Be}$ value for $\pi^-A$ collisions at
 125~GeV/{\it c} is about 4 times smaller than that obtained for $pA$ collisions
 at 800~GeV/{\it c}. Some part of this difference (say, factor $\sim1.5$) can be
 connected with smaller
 pion-nucleon ($\pi N$) cross section in comparison with $NN$ cross section.
 Another part of the difference can be connected with the dependence of nuclear
 effects on the initial energy.

 Unfortunately the errors of the data combined with the analysis errors result
 in too large error of the final result, which
 does not allow us do draw a definite conclusion.

 \section{$A$-dependence of Drell-Yan production at large $x_{\rm F}$}
 Since the Drell-Yan mechanism produces lepton pairs which only interact electromagnetically,
 the $A$-dependence is expected to be weak because no final-state interactions
 affect the lepton pair. However some initial-state interactions may affect the A dependence.

 Almost all the experimental results on Drell-Yan production are presented in terms of
 the ratio of inclusive differential cross sections on a heavy nucleus to that on
 a light one. And there is no opportunity to extract desired spectra separately as was done
 in the previous section.

 However we developed a Monte-Carlo (MC) event generator (HARDPING ---
 Hard Probe Interaction Generator) that extends well known HIJING MC
 \cite{Hijing} on Drell-Yan pair production process and some initial-state
 effects are accounted for
 \cite{Berdnikov1,Berdnikov2}. HARDPING MC describes well
 the data on Drell-Yan pair production in hadron-nucleus ($hA$) collisions
 at high-energies \cite{Berdnikov1,Berdnikov2}.

\begin{figure}[htb]
 \centering
 \includegraphics[width=1.\hsize]{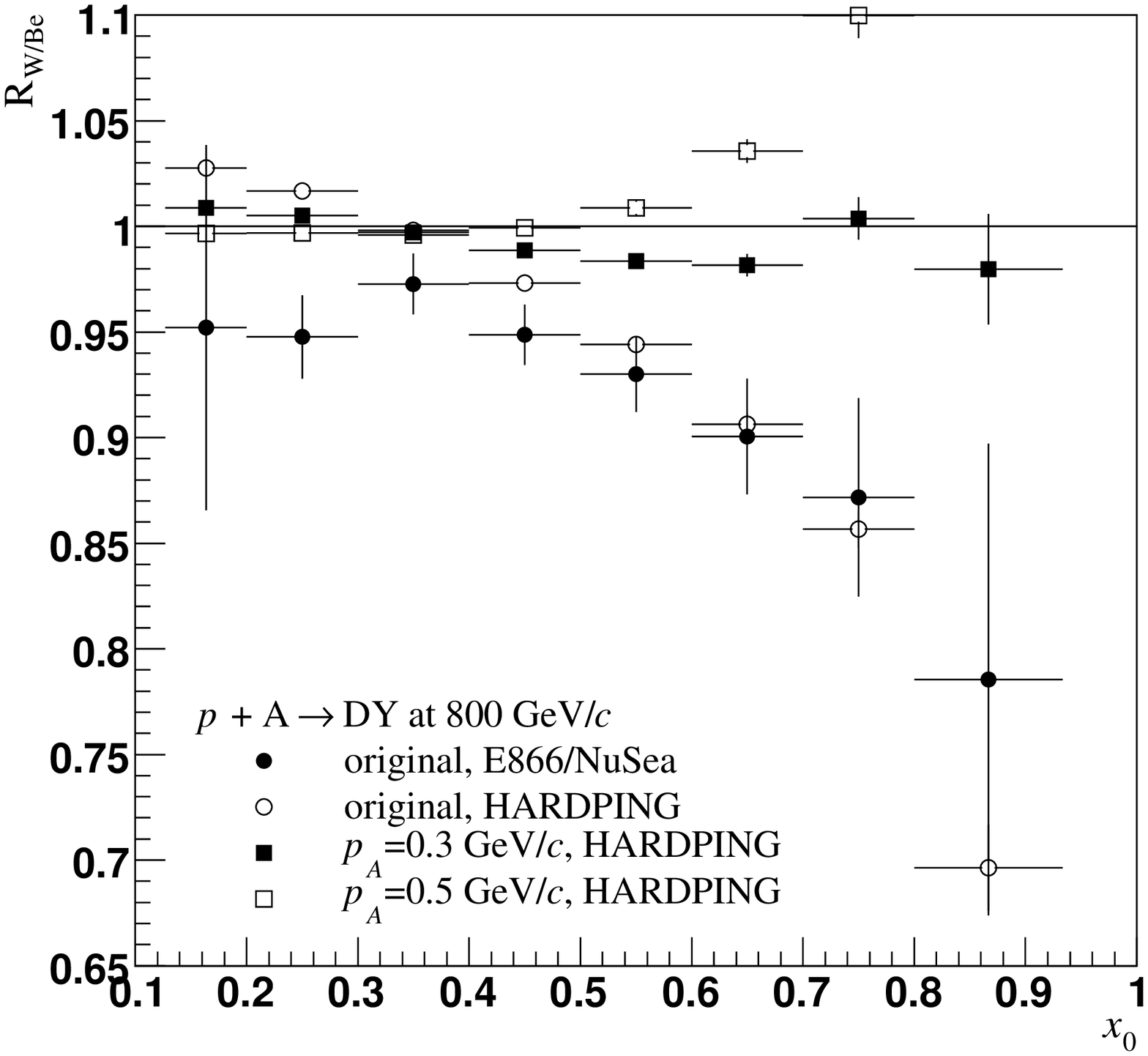}
 \caption{The ratios of inclusive differential cross sections for Drell-Yan
 pair production in $pA$ collisions at 800~GeV/{\it c}, calculated for
 different $p_A$ values.}
 \label{fig:DrellYan}
 \end{figure}

 Using HARDPING MC we simulated absolute spectra for Drell-Yan pair
 production in $pA$ collisions on W and Be targets
 at 800~GeV/{\it c}. Then we applied the procedure described above
 for the two simulated spectra. The results of the analysis are presented
 in Fig.~\ref{fig:DrellYan}. The figure represents the original
 E866 data \cite{Vasiliev1} (solid circles), simulated ratio
 without any shift (open circles) to demonstrate consistency
 between the simulated results and the experimental data,
 and shifted ratios obtained with HARDPING MC. As was predicted
 the fraction of energy/momentum lost by the projectile on
 nuclear effects is small for the case of Drell-Yan pair
 production. This is because there is no final-state interactions
 in Drell-Yan production process.

 The obtained value for $p_A \approx 0.3$~GeV/{\it c} corresponds
 to $\Delta p^{\rm lab.}_{\rm W/Be} \approx 12$~GeV/c.

 \section{Conclusion}
 In summary, we considered the energy/momentum losses of the projectile in $hA$ collisions
 at high-energies from the available experimental data. What we were interested in
 is how much energy/momentum of the projectile is spent on all the nuclear effects
 including the effect of nuclear disintegration.

 The energy losses estimated from the experimental data are in reasonable agreement
 with theoretical calculations \cite{Kop1,Kop2}.
 Namely, our result $\Delta p^{\rm lab.}_{\rm W/Be} \approx 50$~GeV/c
 for $J/\Psi$ production at 800~GeV/c correspond to energy loss rate
 \begin{equation}
 \label{eq:concl_pA_JPsi}
 {\rm d}E/{\rm d}z \approx 5{\rm ~GeV/fm},
 \end{equation}
(We assume the length of full trajectory $\sim 1.5 R_A$ and we neglect
nuclear effects in Be target). The analysis \cite{Kop2} predicts
${\rm d}E/{\rm d}z \approx 3$~GeV/fm for initial-state quark energy loss
rate, i.e. energy loss before the hard interaction point. However
initial-state quark energy losses is not the dominant effect in $J/\Psi$
production processes on nuclear targets. The main contribution to the
suppression of $J/\Psi$'s arises from final-state interactions. Also
there exists strong gluon shadowing at large $x_0$ as well as gluon
enhancement at small $x_0$. Nevertheless, Ref.~\cite{Kop2} does not
present how much projectile energy is spent on each nuclear effect.
(The effects of gluon shadowing for $J/\Psi$ production were calculated
in Ref. \cite{PSS}.) Thus we assume that main part of our result
Eq.~(\ref{eq:concl_pA_JPsi}) is explained by the mentioned effects.

The estimations of energy losses in pairs Be-W and Be-Cu are larger
than the ratio of $A^{1/3}$ values (i.e. length of trajectory) for
W and Cu nuclei. It can be connected with
rather large error bars, or with the $A$-dependence of energy losses more
strong than $A^{1/3}$ behavior. The last reason is not excluded
because the multiplicity of secondary protons produced in hadron--nucleus
collisions in the nuclear targets fragmentation region with energies
$\leq 1$ GeV has $A$-dependence more strong than $A^{1/3}$
\cite{GGC,Fred}. On the other hand, the energies of these protons are
determined by energy losses of the incident particle.

 In the case of Drell-Yan pair production calculations \cite{Kop1}
 predict quark energy loss rate
 ${\rm d}E/{\rm d}z \approx 3$~GeV/fm. There also exists
 shadowing effect in nuclear target, however there is no numerical estimate
for this effect in Ref.~\cite{Kop1}, it was considered in Ref.
\cite{APSS}. Our result $\Delta p^{\rm lab.}_{\rm W/Be} \approx 12$
 ~GeV/c obtained for Drell-Yan production corresponds to energy loss rate
 ${\rm d}E/{\rm d}z \approx 1.2$~GeV/fm for full trajectory.
 However one should take into account the path of the projectile before
the hard interaction point only, which is $\approx 4.4$~fm for W target\cite{Kop1}.
Thus ${\rm d}E/{\rm d}z \approx 2.7$~GeV/fm.
 Assuming that shadowing effect is small
 for Drell-Yan production at 800~GeV/{\it c} we conlude that our results are in
 reasonable agreement with the \cite{Kop1} calculations.

 It is necessary to note that some part of QCD energy losses can be used for
 nuclear destruction and fragmentation due to the final-state interactions.
 This can explain rather large energy of secondary target nucleons and nuclear
 fragments observed in \cite{Azimov1,Azimov2}.

 From the difference in the values of $\Delta p^{\rm lab.}_{\rm W/Be}$
 for $pA$ and $\pi A$ collisions we can conclude that nuclear effects
 probably depend on projectile energy.

\vskip 0.3cm
We are grateful to V. T. Kim, N. N. Nikolaev and M. G. Ryskin for
discussions.

\newpage

 \begin{thebibliography}{99}
 \bibitem{Baldin1} A. M. Baldin {\it et al.}, Yad. Fiz. {\bf 18}, 79 (1973).
 \bibitem{Bayukov1} Yu. D. Bayukov {\it et al.}, Yad. Fiz. {\bf 18}, 1246
(1973).
 \bibitem{Frankfurt1} L. L. Frankfurt and M. L. Strikman, Phys. Rept.
{\bf 76}, 215 (1981).
 \bibitem{Vogt1} R. Vogt, Phys. Rept. {\bf 310}, 197 (1999).
 \bibitem{Vasiliev1} M. A. Vasiliev {\it et al.} (E866 Collab.),
Phys. Rev. Lett. {\bf 83}, 2304 (1999).
 \bibitem{Kop1} M. B. Johnson {\it et al.}, Phys. Rev. {\bf C65}, 025203
(2002).
 \bibitem{Kop2} B. Kopeliovich {\it et al.}, Nucl. Phys. {\bf A696}, 669
(2001).
 \bibitem{Azimov1} S. A. Azimov {\it et al.}, Yad. Fiz. {\bf 8}, 933 (1968).
 \bibitem{Azimov2} S. A. Azimov {\it et al.}, Z. Phys. {\bf A300}, 47 (1981).
 \bibitem{Badier1} J. Badier {\it et al.} (NA3 Collab.), Z. Phys.
{\bf C20}, 101 (1983).
 \bibitem{Katsanevas1} S. Katsanevas {\it et al.} (E537 Collab.),
Phys. Rev. Lett. {\bf 60}, 2121 (1988).
 \bibitem{Leitch1} M. Leitch {\it et al.} (E866 Collab.),
Phys. Rev. Lett. {\bf 84}, 3256 (2000).
 \bibitem{Alde1} D. M. Alde {\it et al.} (E772 Collab.),
Phys. Rev. Lett. {\bf 66}, 133 (1991).
 \bibitem{Kowitt1} M. S. Kowitt {\it et al.} (E605/E789 Collab.),
Phys. Rev. Lett. {\bf 72}, 1318 (1994).
 \bibitem{Gribushin1} A. Gribushin {\it et al.} (E672/E706 Collab.),
Phys. Rev. {\bf D62}, 012001 (2000).
 \bibitem{Abolins1} M. A. Abolins {\it et al.}, Phys. Lett. {\bf B82},
145 (1979).
 \bibitem{Hijing} M. Gyulassy and X.-N. Wang, Comput. Phys. Commun.
{\bf 83}, 307 (1994).
 \bibitem{Berdnikov1} Ya. A. Berdnikov {\it et al.}, Yad. Fiz. (in press).
 \bibitem{Berdnikov2} Ya. A. Berdnikov {\it et al.}, Eur. Phys. J. (in press).
\bibitem{PSS} C. Pajares, C. A. Salgado and Yu. M. Shabelski, Mod.
Phys. Lett. {\bf A13}, 453 (1998).
\bibitem{GGC} K. G. Gulamov, U. G. Gulyamov amd G. M. Chernov, Fiz.
Elem. Chast. Atom. Yadra {\bf 9}, 554 (1978).
\bibitem{Fred} S. Fredriksson {\it et al.}, Phys. Rep. {\bf 144}, 187
(1987).
\bibitem{APSS} N. Armesto {\it et al.}, Yad. Fiz. {\bf 61}, 125 (1998).

 \end{thebibliography}
 \end{document}